\documentclass[12pt]{aastex}
\usepackage{emulateapj5}
\usepackage{rotating}

\newcommand{\kms}{\mbox{\,km\,s$^{-1}$}}
\newcommand{\Kkms}{\mbox{\,K\,km\,s$^{-1}$}}
\newcommand{\Msun}{\,M$_{\odot}$}
\newcommand{\HII}{\mbox{$\mathrm{H\,{\scriptstyle {II}}}$}}
\newcommand{\vlsr}{\mbox{V$_{LSR}$}}
\newcommand{\vsun}{\mbox{V$_{\odot}$}}
\newcommand{\tastar}{\mbox{T$_{A}^{*}$}}
\newcommand{\trstar}{\mbox{T$_{R}^{*}$}}
\newcommand{\tmb}{\mbox{T$_{mb}$}}

\newcommand{\co}{\mbox{$^{12}${\rmfamily CO}{~$J=1\rightarrow 0$}}}
\newcommand{\tco}{\mbox{$^{13}${\rmfamily CO}{~$J=1\rightarrow 0$}}}
\newcommand{\transition}{\mbox{$J=1\rightarrow 0$}}
\newcommand{\cont}{\mbox{$^{12}${\rmfamily CO}}}
\newcommand{\tcont}{\mbox{$^{13}${\rmfamily CO}}}

\slugcomment{}

\shorttitle{The BU-FCRAO GRS}
\shortauthors{Jackson et al.}

\begin{document}

\title{The Boston University--Five College Radio Astronomy Observatory Galactic Ring Survey}

\author{J.~M.~Jackson, J.~M.~Rathborne, R.~Y.~Shah, R.~Simon\altaffilmark{1}, T.~M.~Bania, D.~P.~Clemens, E.~T.~Chambers, A.~M.~Johnson, M.~Dormody, R.~Lavoie}
\affil{Institute for Astrophysical Research, Boston University, 725 Commonwealth Avenue, Boston, MA 02215; jackson@bu.edu, rathborn@bu.edu, ronak@bu.edu, simonr@ph1.uni-koeln.de,  bania@bu.edu, clemens@bu.edu, etc1@bu.edu, alexj@bu.edu, mdormody@bu.edu, coldfury@bu.edu}
\and
\author{M.~H.~Heyer}
\affil{Department of Astronomy, Lederle Graduate Research Tower, University of Massachusetts, Amherst, MA 01003; heyer@astro.umass.edu}

\altaffiltext{1}{{Present address, I.Physikalisches Institut, Universit$\ddot{\mathrm a}$t zu K$\ddot{\mathrm o}$ln, Z$\ddot{\mathrm u}$lpicher Strasse 77, 50937 K$\ddot{\mathrm o}$ln, Germany.}}

\begin{abstract}
The Boston University-Five College Radio Astronomy Observatory Galactic Ring Survey
is a new survey of Galactic \tco\, emission. The survey used the SEQUOIA multi pixel array on the 
Five College Radio Astronomy Observatory 14 m telescope to cover a longitude range of 
$\ell =$ 18$\arcdeg$--55.7$\arcdeg$ and a latitude range of $|b| < 1\arcdeg$, a total of 
75.4 square degrees. Using both position-switching and On-The-Fly mapping modes, we achieved 
an angular sampling of 22$\arcsec$, better than half of the telescope's 46$\arcsec$ 
angular resolution. The survey's velocity coverage is $-$5 to 135\,\kms\,
for Galactic longitudes $\ell \leq 40\arcdeg$ and $-$5 to 85\,\kms\, for Galactic 
longitudes $\ell > 40\arcdeg$.
At the velocity resolution of  0.21\,\kms, the typical rms sensitivity is 
$\sigma$(\tastar) $\sim$ 0.13\,K.
The survey comprises a total of 1,993,522 spectra. We show integrated intensity 
images (zeroth moment maps), channel
maps, position-velocity diagrams, and an average spectrum of the completed survey dataset.  We
also discuss the telescope and instrumental parameters, the observing modes, the data reduction 
processes, and the emission and noise characteristics of the dataset. The Galactic Ring Survey data are available to the 
community at www.bu.edu/galacticring or in DVD form by request.
\end{abstract}


\section{Introduction}
The Galactic distribution of molecular hydrogen was first deduced from early CO observations 
\citep{Burton75,Scoville75}. Surprisingly, the CO had a Galactocentric radial distribution 
distinct from that of atomic hydrogen. In particular, the CO distribution in the northern hemisphere 
showed a large peak in the molecular gas density about midway between the Sun and the Galactic Center. Subsequent 
determinations of the face-on distribution of CO showed that this feature, dubbed the ``5 kpc ring,'' dominates the 
Galaxy's molecular gas structure \citep{Clemens88}. 

Most of the Galaxy's star formation activity takes place in the 5 kpc ring.  With a mass of 
2$\times$10$^{9}$\Msun, the 5 kpc ring contains about 70\% of all of the molecular gas 
inside the solar circle \citep{Clemens88}. The ring is thus an enormous reservoir of material 
for the formation of new stars and clusters. Indeed, most of the Galactic giant \HII\, regions, 
far infrared luminosity, diffuse ionized gas, and supernova remnants are associated with 
the ring \citep{Burton76,Robinson84}.   

Because the 5 kpc ring dominates both the molecular interstellar medium (ISM) and the star formation activity 
in the Milky Way, it plays a crucial role in the dynamics, structure, and chemical 
evolution of our Galaxy. Yet, despite its obvious importance, the 5 kpc ring has remained poorly 
understood, mostly due to the challenge of imaging such a large 
expanse of sky (several tens of square degrees) at good angular resolution and sensitivity.

Recent advances in millimeter-wave array technology now allow a comprehensive study of molecular gas 
in the 5 kpc ring. In particular, the development of monolithic microwave
integrated circuits (MMICs) has enabled the construction of large focal
plane array receivers at millimeter wavelengths and a corresponding
improvement in mapping speed.  Using such an array, we have
conducted a new \tco\, molecular line survey of the inner Galaxy, the Boston University-Five 
College Radio Astronomy Observatory Galactic Ring Survey (GRS). 

Although many previous surveys mapped the Galaxy in the main isotopic \co\, transition, we 
have chosen to conduct the GRS in the \tco\, transition. Because \tcont\, is much 
less abundant than \cont, the \tcont\, transition
has a smaller optical depth.  It is therefore a better column density
tracer than \cont.  Moreover, the smaller optical depths result in
narrower linewidths for \tcont, and consequently, blended lines from two
distinct clouds at similar velocities can be separated more cleanly.

In this paper, we describe the survey, the data reduction procedures, and the
emission and noise characteristics of the dataset. Analyses of the survey will be
published in future papers.

\section{The Survey}

\subsection{Telescope, Receiver, and Backend Parameters}

The GRS was conducted using the Five College Radio Astronomy Observatory (FCRAO) 14~m telescope 
located in New Salem, Massachusetts between 1998 December and 2005 March. The survey covers the region between 
18$\arcdeg<\ell<55.7\arcdeg$ and $|b|<1\arcdeg$. At the \tco\, frequency ($\nu_0=110.2~{\rm GHz}$),
the beam-width of the FCRAO 14~m is 46\arcsec. 

All observations were obtained using the single sideband focal plane array
receiver SEQUOIA (SEcond QUabbin Optical Imaging Array; \citealp{Erickson99}).  SEQUOIA
employs low-noise MMIC based amplifiers.  Before 1999, the MMICs were InP which
provided a mean receiver noise temperature of 80 K. After 1999, SEQUOIA was
upgraded to InSb MMICs which provided a mean receiver noise temperature of 60 K. 

For the observing period 1998--2001, SEQUOIA contained 16 separate receiving elements.
These elements were arranged in a singly-polarized 4$\times$4 array with a
separation between elements of 88\arcsec\, on the sky. In this observing period,
SEQUOIA was used in combination with a spectrometer consisting of 16
autocorrelators. Each autocorrelator had a bandwidth of 40 MHz, 512 channels,
and a velocity sampling of 0.21\,\kms. The autocorrelators were centered at a
velocity of 40\,\kms\, and thus covered the velocity range $-$10 to 90\,\kms.
This setup was sufficient to cover the local standard of rest (LSR) 
velocities\footnote{To deduce our LSR velocity scale we used the IAU J2000 standards for the solar
motion with respect to the LSR of $\alpha$ = 271\arcdeg\, and $\delta$=30\arcdeg. 
For consistency with previous FCRAO observations we use \vsun=20\kms\, rather than the
IAU standard of \vsun=19.4\kms.} for most of the \tcont\, emission for Galactic 
longitudes $\ell >$ 40$\arcdeg$.

In 2002 January, a new autocorrelator, the dual channel correlator (DCC), was
commissioned. The DCC enabled the processing of two independent intermediate
frequencies (IFs). The DCC was configured to have a bandwidth of 50 MHz, 1024
channels, and a velocity sampling of 0.13\,\kms\, for both of the IFs. In order
to provide enough velocity coverage to sample all Galactic \tcont\, emission
below Galactic longitudes of 40$\arcdeg$, the two IFs were both tuned to the \tcont\,
transition but centered separately at LSR velocities of 20\,\kms\, and 100\,\kms.  
This setup provided 270 channels of overlap between the IFs and thus allowed us to 
generate a composite spectrum for each position on the sky with a total of 210\,\kms\, 
bandwidth. In 2002 March, SEQUOIA was upgraded
from a 16 element array to a 32 element dual-polarization array with two IFs for
each of the 32 detector elements.  The combination of the upgraded SEQUOIA and
the DCC enabled processing of both IFs for each of the 32 pixels, resulting in
64 simultaneously observed spectra.

\subsection{Pointing Checks}

Pointing checks were conducted at the beginning of every observing session
(typically lasting a total of eight hours) using the
SiO masers T~Ceph or $\chi$~Cyg. When the observing session extended beyond sunrise (April--May),
additional pointing checks were also performed using the SiO maser R~Cas. 

The pointing offsets were determined using a five-point pattern and were 
adjusted iteratively to find the optimum pointing of the telescope.
When the fits to the pointing correction coverged to $\le$ 5\arcsec\, no more
pointing corrections were adopted. At this point, the focus was then adjusted
and a parabolic fit to the resultant intensities determined the optimal focus. 
Once the optimal focus was achieved, the five-point pointing procedure was repeated.
We therefore estimate that the pointing accuracy is better than 5\arcsec.

\subsection{Selection of Emission-Free Sky Positions}

To account for millimeter wave emission from the sky, it is necessary to
subtract the sky emission from a nearby ``off'' position from the spectrum of
the source position. Because the Galactic plane contains extended, strong
\tcont\, emission, we had to choose ``off'' positions carefully to avoid
contaminating the source position with unwanted emission in the ``off'' position.

To select ``off'' positions with little or no \tcont\, emission, we examined \cont\, maps
from the Columbia/CfA survey in order to find positions as close as possible 
to the Galactic plane with little or no emission (T.~M.~Dame 1998, private communication). We then searched these 
potential emission-free positions in \tcont\, by position-switched observations against 
regions from the Columbia/CfA survey known to be free of emission in \cont\, to an rms noise 
level of 0.05 K (for 0.65\,\kms\,velocity resolution). For all but one position, no \tcont\, emission 
was found toward the sky positions above a noise level of \tmb=0.07~K. One of the ``off'' 
positions, however, possesses a narrow \tcont\, emission feature \tastar\, $\sim$ 0.1~K at \vlsr\,=15\,\kms, 
affecting data in the range $\ell$=32$\fdg$5 to 36\arcdeg\, and $b$=$-$0.5 to 0$\fdg$8.  
This contamination manifests itself as a negative spectral artifact feature in the final dataset. 
In total fifty-two ``off'' positions were used for the GRS. For each map we selected the
nearest possible ``off'' position, which was typically $<$ 2\arcdeg\, from the
source position in both Galactic longitude and latitude.

\subsection{Observing Modes}

The majority of the observations were conducted remotely. Due to weather constraints, the 
observing season was typically limited to the months of November through May. From the FCRAO, the 
lowest Galactic longitude covered by the GRS reached a maximum elevation of about 45$\arcdeg$. 
To avoid observations at high air mass, all GRS data were obtained at elevations $>$29$\arcdeg$. 
To maintain uniform data quality, we did not observe if the system temperature exceeded
400~K. The typical system noise temperatures were 200--300~K.

We employed three separate observing modes to map the Galactic
\tco\, emission: position-switching with the 16-element 
SEQUOIA (1998 December--2001 May), On-The-Fly (OTF) mapping with the 16-element SEQUOIA 
(2002 Jan--March), and OTF mapping with the upgraded 32-element SEQUOIA (2002 March--2005 March). 

In the position-switching mode, the data were obtained by stepping the array
$22\farcs 14$ in a 16$\times$16 grid (these grids were referred to as ``footprint''). In 
this mode the dewar containing the focal plane array receiver was rotated
continuously to align the array in Galactic coordinates.  This scheme produced a
fully sampled ``footprint'' map with a size of $\ell \times b$ = 5$\farcm 9
\times 5\farcm 9$. Four consecutive array pointings shared the same emission-free
``off'' position.  Integration times per point for one on-source pointing ranged
between 10 and 20~s, with 20 to 40~s spent on the emission-free ``off'' position.
Most positions were observed twice to achieve the desired sensitivity.  The
resulting spectra have an average 1 $\sigma$ rms noise level \tastar\,= 0.20~K
per 0.21\kms\, channel. In a typical eight-hour observing session, we covered 
$\sim$ 0.35 deg$^{2}$. Due to the limitations in the velocity coverage of the
autocorrelators, the observations during this period were restricted to Galactic
longitudes $>$40$\arcdeg$, where the velocity of the Galactic \tcont\,
emission is always $<$85\,\kms.  In the 2.5 years using
this mode, we mapped the region from $\ell$ = 40$\arcdeg$ to 51$\arcdeg$ and $b$ = $-$1$\arcdeg$ to
0.5$\arcdeg$ (16 deg$^{2}$; 21.2\% of the entire survey).

In 2002 January, the OTF mapping mode was implemented on the FCRAO. In this mode, the 
telescope is scanned across the sky and data are collected
as the telescope moves.  Because the data storage, slew, and settling times are greatly
reduced, data obtained in this mode have far less overhead and are consequently more efficiently collected.
   Since in the OTF mode the SEQUOIA dewar does not rotate  to compensate
  for the field rotation, the mapped region is covered on a highly irregular,
  yet very densely-sampled grid. This has several consequences. In the central
  region of the OTF map, every pixel is scanned over every position in the map.
  Compared to the position-switching mode, this redundancy enhances the data
  quality since pixel-to-pixel variations are averaged. Artifacts due to gain  and noise
  variations among pixels, therefore, are greatly reduced. Towards the map
  edges, the coverage becomes less dense. Therefore, the noise is only uniform
  in the inner, well sampled part of the map and increases towards the map
  edges. To compensate for this effect and achieve more uniform noise in the final
  data set, we overlapped many individual maps to cover the survey region. 
The survey region in OTF mode was broken up into blocks 
of $\ell$$\times$$b$ = 6\arcmin$\times$18\arcmin. Each OTF map consisted 
of a single one of these blocks. For each map, the telescope scanned 
in Galactic latitude at a rate of 0.35\,arcmin s$^{-1}$ in a raster pattern. 
To sample the emission within the mapping region sufficiently, the array was offset 
by 0.75 of the beam (35\arcsec) between scanning rows. 
The data were read out at a frequency of 1 Hz as the telescope
scanned so that 1 sec integration time was spent on each of the 64 readouts
(32 array elements, two velocity settings). We integrated for 10 sec on the 
emission-free ``off'' position, after every second row ($\sim$ every 2 mins).

Because we scanned the array during observations, the beam may be
slightly elongated for the OTF data. This will only affect the beam in the 
scanning direction and is estimated to be $<$ 5 \% ($< 2\farcs3$).

If weather permitted, in an eight-hour observing session we completed $\sim$ 20 OTF maps
($\sim$ 0.6 deg$^{2}$). Within each observing session, the maps were chosen to span specific ranges
in Galactic longitude depending on the local sidereal time, so that we were always obtaining 
data at high elevations and hence low system temperatures. 

In the three months of using the OTF mode with the 16-element SEQUOIA, we
surveyed a total solid angle of 5.3 deg$^{2}$ (7.0 \% of the entire survey).
The spectra obtained in this mode have a mean 1 $\sigma$ rms noise level
\tastar\,= 0.18~K per 0.21\,\kms\, channel. From March 2002, we used the
upgraded 32-element SEQUOIA in the OTF mode.  We completed the remainder of the
survey region in this mode, $\sim$ 54.1 deg$^{2}$ (71.8 \% of the entire
survey).  These spectra have a mean 1 $\sigma$ rms noise level \tastar\,= 0.13~K
per 0.21\,\kms\, channel. For a more complete discussion of the noise see
$\S$ \ref{noise}.

\subsection{Flux Calibration}

The spectra were calibrated using a vane to switch between emission from the sky and an ambient 
temperature load. The calibration occurred before the start of each map and after every 10
rows during the map ($\sim$ every 10 mins). This
method of calibration produces typical errors in the temperature scale of 
10--15\%.

To verify the system setup and to keep a record of the system
performance over the duration of the survey, we obtained a single position-switched 
observation of W51 at the start of each observing session.
An analysis of these data show that the W51 peak brightness 
temperature showed an rms dispersion of $\sim$ 15\%.  We use this value as 
an estimate of the accuracy of our flux scale.

All intensities are reported on the antenna temperature scale, 
\tastar\, (K). To convert  antenna temperatures to main-beam brightness temperatures, 
the \tastar\, values should be divided by the main-beam 
efficiency $\eta_{mb}$ of 0.48 derived from observations of planets.

\subsection{Data Reduction}

The position-switched raw data were converted to CLASS\footnote{CLASS is part of the GILDAS
package.} format. Subsequent analysis was made with the CLASS software package. Each 
``footprint'' was processed individually. To facilitate baseline removal, an average 
spectrum of each ``footprint'' was inspected to determine velocity ranges with
significant emission to be excluded from the polynomial baseline fit. Together with the order of the
baseline fit, these velocity ``windows'' were altered until a suitable fit was achieved.  
The polynomial order of the final baseline fit was typically first (linear) or second (parabolic). 
The velocity windows and the baseline order were then applied to fit spectral baselines for each individual
spectrum within the ``footprint.''

The OTF data were processed at FCRAO with the GUI based program OTFTOOL\footnote{OTFTOOL was written 
by M.H. Heyer, G. Narayanan, and M. Brewer.}. We used this 
program to regrid the irregularly sampled OTF data onto the same $22\farcs 14$ grid as the 
position-switched data. Each raw OTF spectrum was convolved onto a regular grid using
a spatial kernel that minimized aliased noise power while retaining 
the full resolution of the telescope. The kernel is 
\begin{equation}
\frac{J_1(2{\pi}ax)}{2{\pi}ax} \frac{J_1(cx/R_{max})}{cx/R_{max}} e^{-(2bx)^2} \Pi(R_{max}) 
\end{equation}
 and is similar to kernels used at other telescopes 
that have implemented OTF mapping \citep{Mangum00}. $J_1$ is the first order Bessel function.
The variable $x$ is the 
distance from the observed data point from the output grid cell position in units 
of $\lambda/D$ where $\lambda$ is the observed wavelength of the observation and 
$D$ is the diameter of the FCRAO telescope.  $R_{max}$ is the truncation 
radius of the kernel in units of $\lambda/D$,  beyond which the 
kernel is zero.  For all regridding of the GRS data, $R_{max}$= 3.
The coefficients, $a$, $b$, are determined to 
minimize the aliased noise power given the edge taper of the FCRAO telescope.
The values are $a$=0.9009, $b$=0.21505. The coefficient, $c$=3.831706, corresponds 
to the first null of $J_1(x)$. The function $\Pi(R_{max})$ acts as a pill box function
where $\Pi$=1 for R$<$R$_{max}$ and $\Pi$=0 for R$>$R$_{max}$.

The resulting spectra for each IF were converted separately and stored in a single
CLASS spectral file. The next step in the data reduction process was to combine the spectra
from the two IFs. This was achieved by resampling the spectra on a common
velocity grid and producing a composite spectrum by merging the spectra
for the two IFs. Data from $-$10\,\kms\, and 80\,\kms\, from the first IF were merged with data from
45\,\kms\, to 140\,\kms\, from the second IF. For the OTF data, an average spectrum for  
a 0$\fdg$3$\times$0$\fdg$3 region was used to determine the velocity windows and 
baseline order (again, typically linear or parabolic), which were then applied to fit spectral
baselines for each individual merged spectrum within the region.

The OTF data obtained with the higher velocity resolution (0.13\,\kms)
were resampled to 0.21\,\kms\, to match spectra from the position-switching mode.
All spectra were then assembled into 3-dimensional 
data cubes ($\ell$,$b$,$v$) using CLASS routines. These cubes were then converted 
into FITS format using routines within the GILDAS\footnote{The GILDAS working group 
is a collaborative project of the Observatoire de Grenoble and 
Institut de Radio Astronomie Millimetrique (IRAM), and comprises: G. Buisson, L. Desbats, G. Duvert, 
T. Forveille, R. Gras, S. Guilloteau, R. Lucas, and P. Valiron.}
package. Routines written using the CFITSIO libraries were then used to 
merge the individual FITS cubes into the larger cubes that comprise the final data release.

The final data release contains twenty FITS cubes.  Eighteen of these cubes are
$\ell \times b$ = 2$\arcdeg \times 2\arcdeg$, one is $\ell \times b$ = 1$\arcdeg \times 2\arcdeg$,
and one is $\ell \times b$ = 0$\fdg7 \times 2\arcdeg$. The 2$\arcdeg \times 2\arcdeg$ data cubes 
are centered at
even integral values of Galactic longitudes from 20$\arcdeg$\, to 54$\arcdeg$. The 1$\arcdeg \times 2\arcdeg$
cube is centered at a Galactic longitude of 18$\fdg$5, and the 0$\fdg7 \times 2\arcdeg$ cube is centered at 
a Galactic longitude of 55$\fdg$35. All cubes are
centered in Galactic latitude at $b$ = 0$\arcdeg$. Cubes centered at Galactic
longitudes $\leq$ 40$\arcdeg$ span a velocity range of \vlsr\, = $-$5 to 135\,\kms,
while cubes centered at Galactic longitudes $>$ 40$\arcdeg$ span a velocity range 
of \vlsr\,=$-$5 to 85\,\kms (in both cases the data within 5\,\kms\, from the edge of the
passband were discarded). All cubes have identical channel
widths of 0.21\,\kms, but a different number of channels depending on whether the center Galactic 
longitude is $\leq$ 40$\arcdeg$ (659 channels) or $>$ 40$\arcdeg$ (424 channels). Although the cube
centered at $\ell$ = 40$\arcdeg$ contains 659 velocity channels, for the region where 
40$\arcdeg < \ell < 41\arcdeg$ 
the velocity channels above 85\,\kms\, are blank. The complete survey contains $\sim$ 5 Gbytes of data.

\section{Comparison to Previous CO Line Surveys}

The GRS has mapped the Galactic plane in the \tco\, transition.  Unlike the optically thick \co\, 
transition, which suffers from velocity crowding, the \tco\, transition has a lower optical depth and, 
therefore, narrower linewidths. \tcont\,
thus allows both a better determination of column density and also a cleaner separation 
of velocity components.  Moreover, because of the improvement in mapping speed allowed 
by an array receiver, the GRS achieves half-beam sampling. Unlike most previous surveys 
of molecular emission, the GRS realizes the full resolution of the telescope. 
Table~\ref{table-survey-comparison} compares the GRS to the previous 
CO surveys: the University of Massachusetts-Stony Brook survey (UMSB; \citealp{Sanders86}), the Bell Labs Survey 
\citep{Lee01}, and the Columbia/CfA survey \citep{Dame87}.

\section{The Data}

\subsection{Integrated Intensity Image}

\begin{sidewaysfigure}
\includegraphics[angle=-90,width=\textwidth]{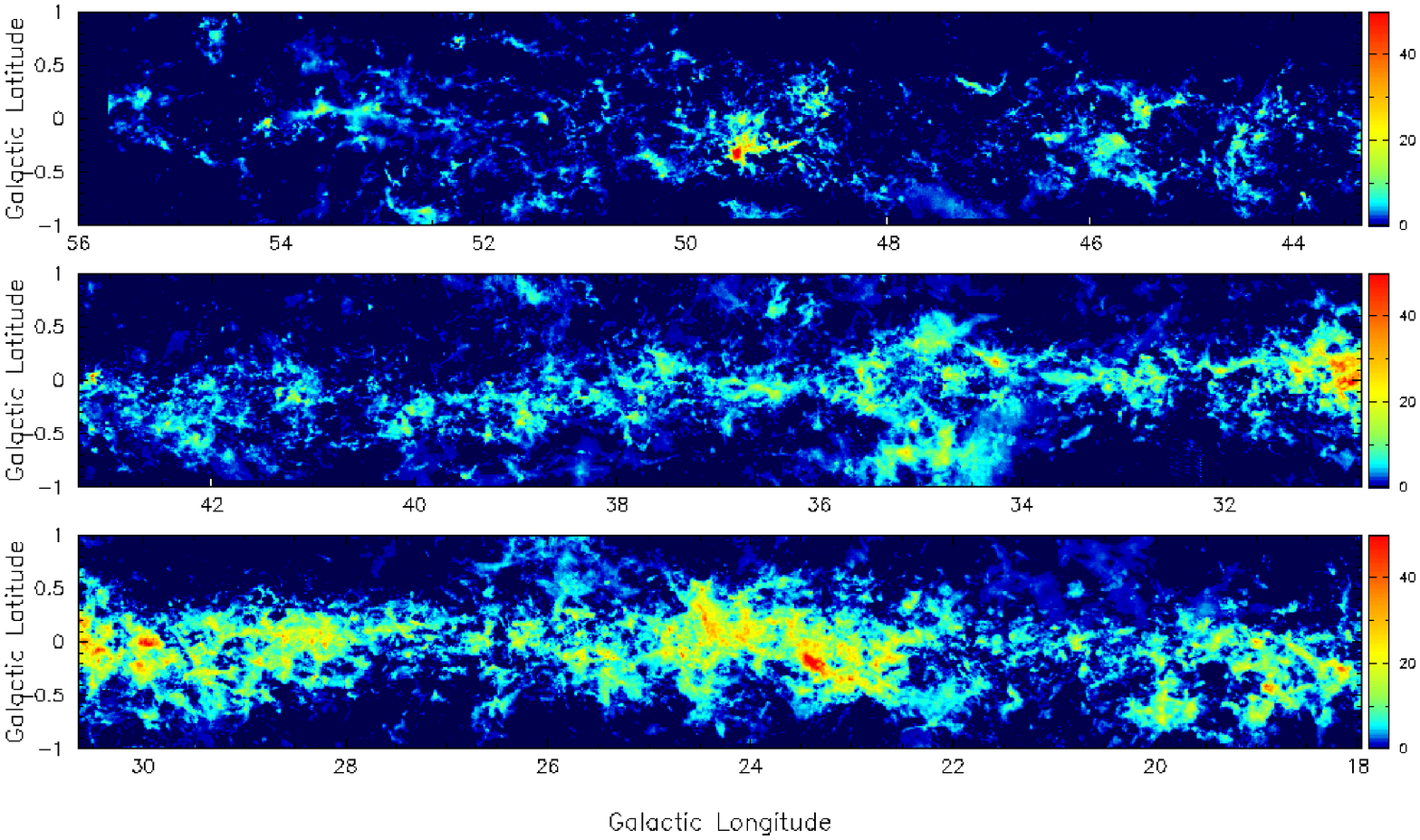}
\caption{\label{grs-momentmaps}Integrated intensity image (zeroth-moment map) of GRS \tcont\, emission 
integrated over all velocities (\vlsr = $-$5 to 135\,\kms\, for Galactic longitudes $\ell \leq 40\arcdeg$ and \vlsr\,= 
$-$5 to 85\,\kms\, for Galactic longitudes $\ell > 40\arcdeg$).
The image shows that most of the emission is confined to $b \sim 0\arcdeg$, with
concentrations at $\ell$~$\sim$~23$\arcdeg$ and $\sim$ 31$\arcdeg$. A 
striking aspect of the image is the abundance of filamentary and linear structures and
the complex morphology of individual clouds. The image is in units of \Kkms.}
\end{sidewaysfigure}

Figure~\ref{grs-momentmaps} shows the integrated intensity image (zeroth moment
map) of GRS \tcont\, emission integrated over all velocities (from \vlsr\,=$-$5 to 135\,\kms\, for
Galactic longitudes $\ell \leq 40\arcdeg$ and \vlsr\,= $-$5 to 85\,\kms\, for 
Galactic longitudes $\ell > 40\arcdeg$).

The map was constructed using a masked moment procedure. In this method, the
intensity integrated over the specified velocity range towards each position
of the map is not simply determined from the sum over all velocity channels.
Instead, for any given position (x,y) and velocity channel i, the
value T$_{i}$dv for that position is only included in the sum of the total
integrated intensity if two conditions are met: (1) the value of both the central pixel
and all 8 neighboring spatial pixels in the same velocity channel exceed the temperature
threshold, and (2) the value of the central pixel and the two neighboring velocity pixels in
adjacent channels also exceed the temperature threshold. 
Details of the method are described by \cite{Adler92}. For the GRS data, we
used a temperature threshold of \tastar\, = 0.6 K.

This image reveals that most of the \tcont\, molecular line emission is 
confined near $b\sim 0\arcdeg$ with concentrations at $\ell$~$\sim$~23$\arcdeg$ and $\sim$ 31$\arcdeg$ 
(peak integrated intensities of 72 and 68 \Kkms\, respectively).
A large number of distinct clouds can be seen in the integrated intensity image.  
The brightest of these correspond to well-known star-forming regions such as W51 and W49.  
A striking aspect of the image is the abundance of filamentary and linear structures and
the complex morphology of individual clouds.

\subsection{Channel Maps}

\begin{sidewaysfigure}
\includegraphics[angle=270,width=0.9\textwidth]{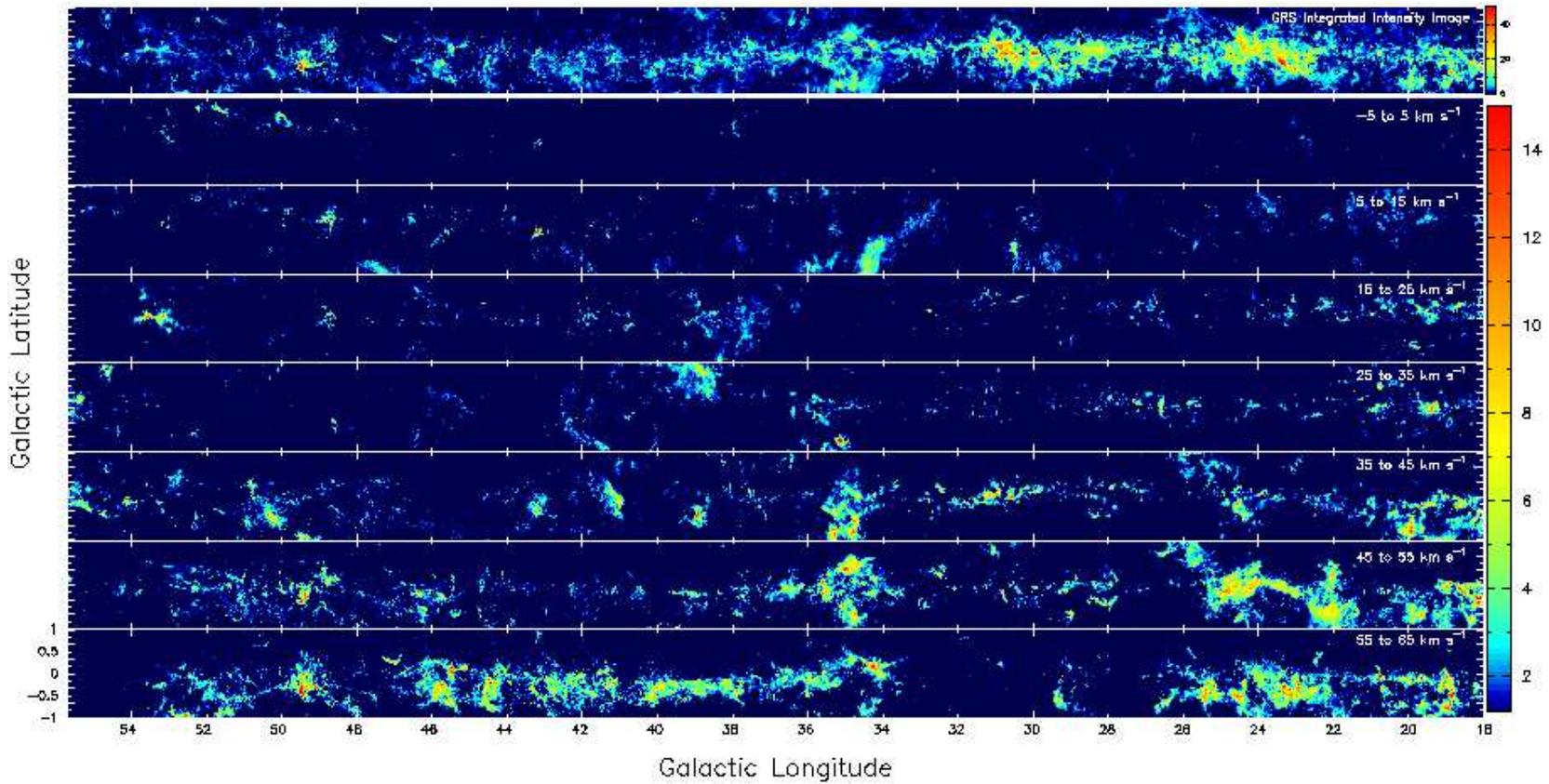}
\caption{\label{channel-maps}Channel maps of GRS \tcont\, emission. The maps were made by
integrating the emission over 10\,\kms\, velocity bins. The image units are in \Kkms. The top image shows
the integrated intensity image from Figure~\ref{grs-momentmaps} for comparison. The velocity
range of each image is labeled. The channel maps clearly separate emission features along the 
same line of sight.}
\end{sidewaysfigure}

Figure~\ref{channel-maps} shows channel maps that were made by integrating the 
\tcont\, emission over 10\,\kms\, velocity bins. The channel maps separate emission 
features along the same line of sight into individual clouds.
In addition, individual clouds too faint to be prominent in the integrated intensity image
often appear as obvious distinct features in the channel maps. 

\subsection{Position-Velocity Diagram}

\begin{sidewaysfigure}
\includegraphics[angle=270,width=0.9\textwidth]{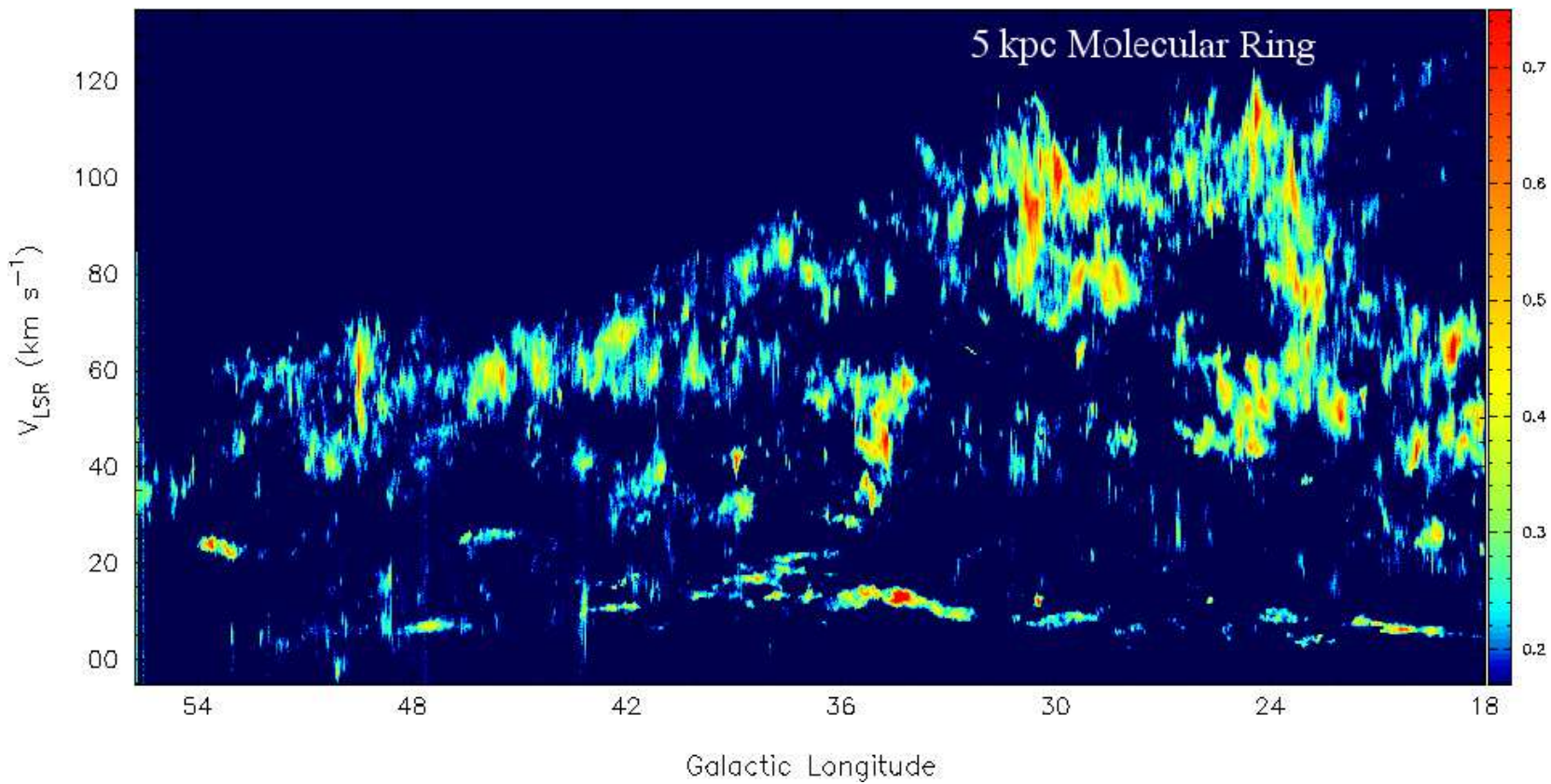}
\caption{\label{lvd}Position-velocity ($\ell$-v) diagram of GRS \tcont\, emission, made by averaging
over all Galactic latitudes ($|b| < 1\arcdeg$). Clearly evident
are the large-scale Galactic features especially the molecular ring (the approximate location
is marked). The image units are in K.}
\end{sidewaysfigure}

\begin{figure}
\centering
\includegraphics[angle=0,height=0.9\textheight]{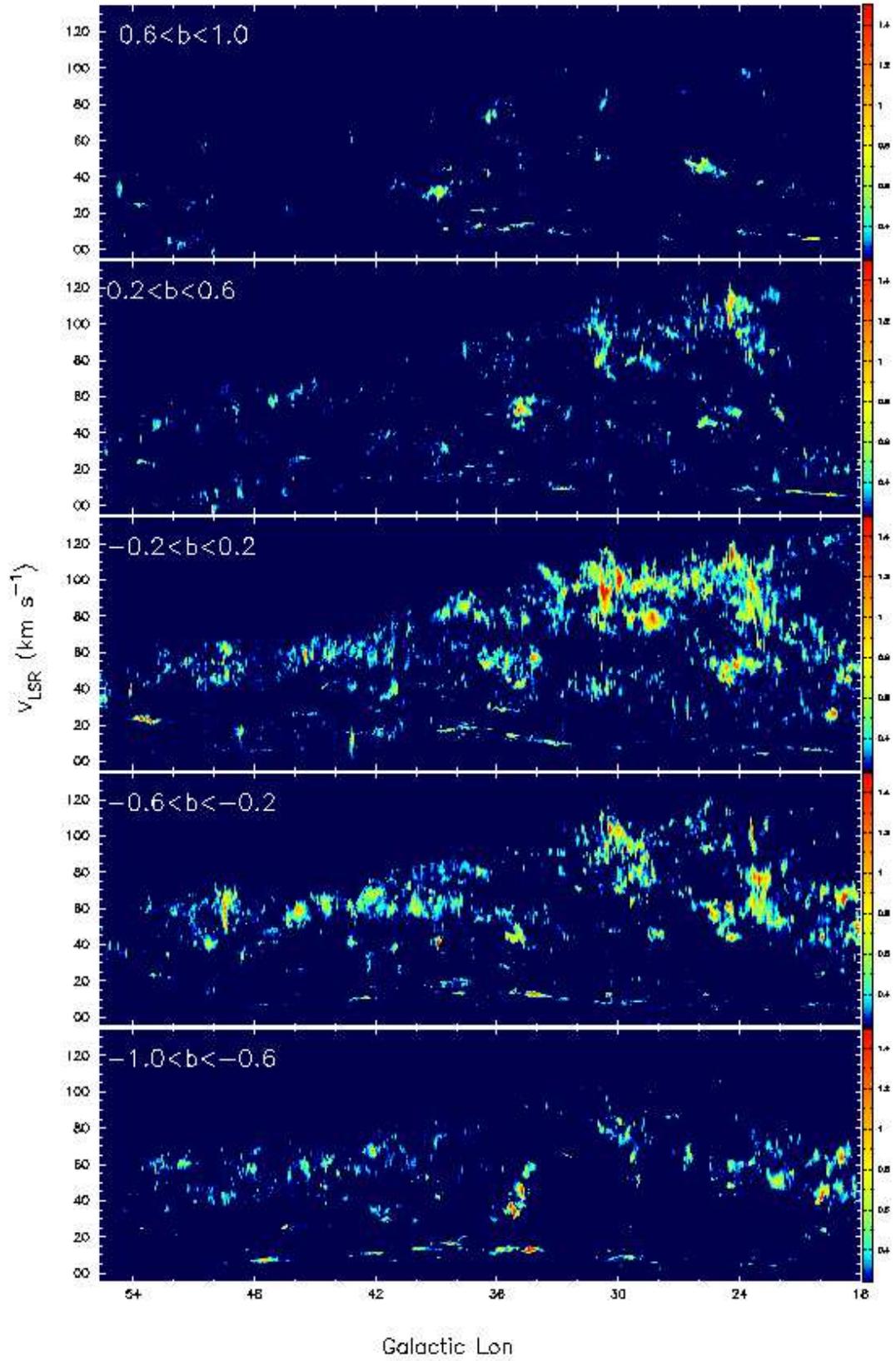}
\caption{\label{lv-cuts}Position-velocity ($\ell$-v) diagram of GRS \tcont\, emission averaged over five ranges
in Galactic latitude. The latitude coverage of each image is labeled. The image units are in K.}
\end{figure}

Figure~\ref{lvd} shows the longitude-velocity ($\ell-$v) diagram of the GRS \tcont\, emission. This
diagram was made  by averaging the emission over the full coverage in Galactic 
latitude. To show how the emission varies as a function of Galactic
latitude, we have also made ($\ell-$v) diagrams over five ranges
in latitude in steps of $\Delta b$ = 0$\fdg$4 (Fig.~\ref{lv-cuts}). 

These position-velocity diagrams reveal large-scale features, especially the feature
commonly called the ``molecular ring.''  We stress that
bright regions in the ($\ell-$v) diagram arise from a complicated
combination of the Galactic column density and the Galaxy's
velocity field, and that the ``molecular ring'' may not 
represent a real ring-like feature.

\subsection{Integrated Spectrum}

\begin{figure}
\epsscale{.80}
\plotone{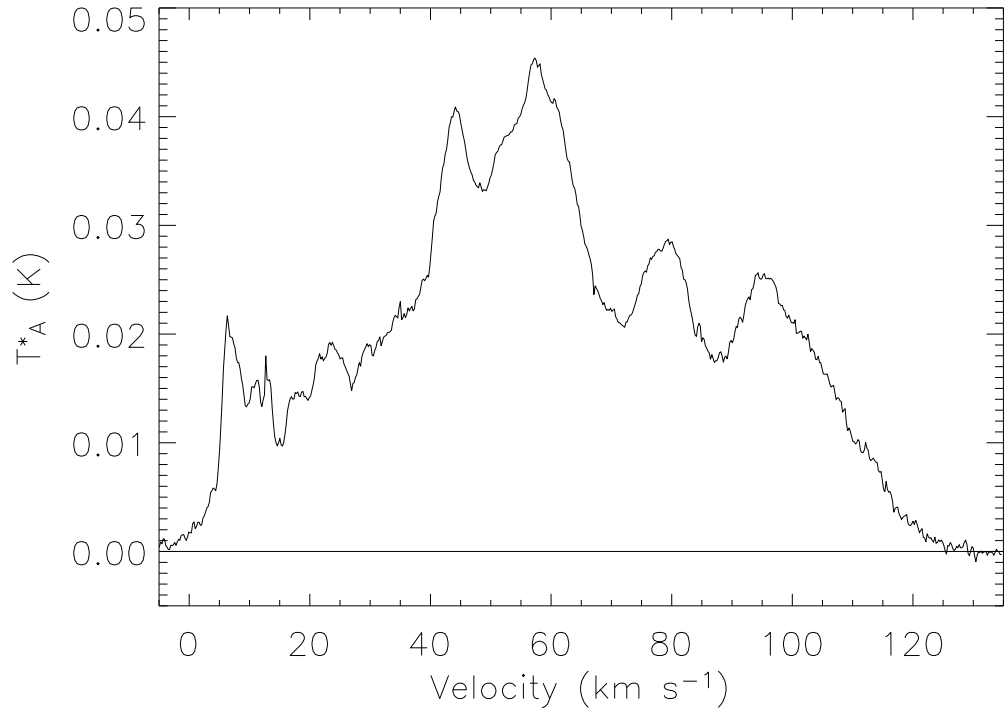}
\caption{\label{grs-spectrum}Average spectrum for the entire GRS. We see \tcont\, emission at all
positive velocities allowed by Galactic rotation, with many distinct peaks.}
\end{figure}

Figure~\ref{grs-spectrum} shows the averaged \tcont\, spectrum for the entire GRS.  
Every position was averaged together with equal weighting to create this 
spectrum.  The effect is somewhat 
analogous to observing a portion of an external edge-on galaxy's disk with a single pixel.  

The GRS detects \tcont\, emission at every positive velocity allowed by Galactic rotation.  
Nevertheless, distinct velocity features can still be recognized, for instance, the peak of the
emission occurs at a velocity of about 57 \kms. Whether these velocity features correspond
to distinct physical structures or merely result from velocity crowding effets is the
subject of future papers.

\section{Data Characteristics}

\subsection{Emission Characteristics}

\begin{figure}
\includegraphics[angle=90,width=0.5\textwidth,clip=true]{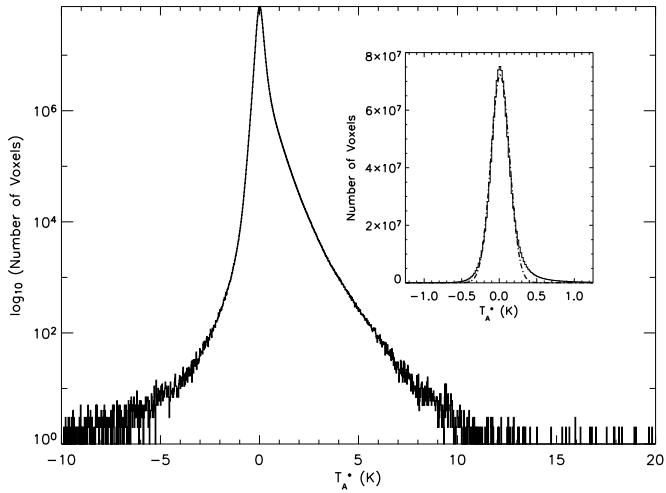}
\caption{\label{emission_stats} Antenna temperature distribution of all ($\ell$,$b$,v) voxels in the GRS. 
The inset shows the same distribution but on a  linear  rather than logarithmic scale. The dash-dotted line 
is a Gaussian fit to the distribution (peak at 0.014 K and full-width at half-maximum of 0.29 K).}
\end{figure}

Figure~\ref{emission_stats} plots a histogram of the antenna temperature distribution 
for all independent ($\ell$,$b$,v) elements (``voxels'') in the GRS. 
The inset shows the same plot but on a linear rather than logarithmic scale.
We also show a Gaussian fit to the distribution in the linear scale plot. The fitted Gaussian
peaks at 0.014~K and has a full-width at half-maximum of 0.29~K.

Purely random noise would produce a voxel distribution centered exactly at zero
with a Gaussian shape. Although a Gaussian fits the data fairly well, there is excess emission, especially in the wings
of the distribution. The positive excess reflects the presence of \tcont\, emission.
The negative excess shows that a single noise
temperature does not characterize the entire dataset. This may be due to variations in 
weather and elevation, different observing modes, weak emission in our ``off'' positions, the poor spatial
sampling at the edge of the OTF maps, or additional systematic errors. 

\subsection{Noise characteristics}
\label{noise}

\begin{sidewaysfigure}
\includegraphics[angle=270,width=0.9\textwidth]{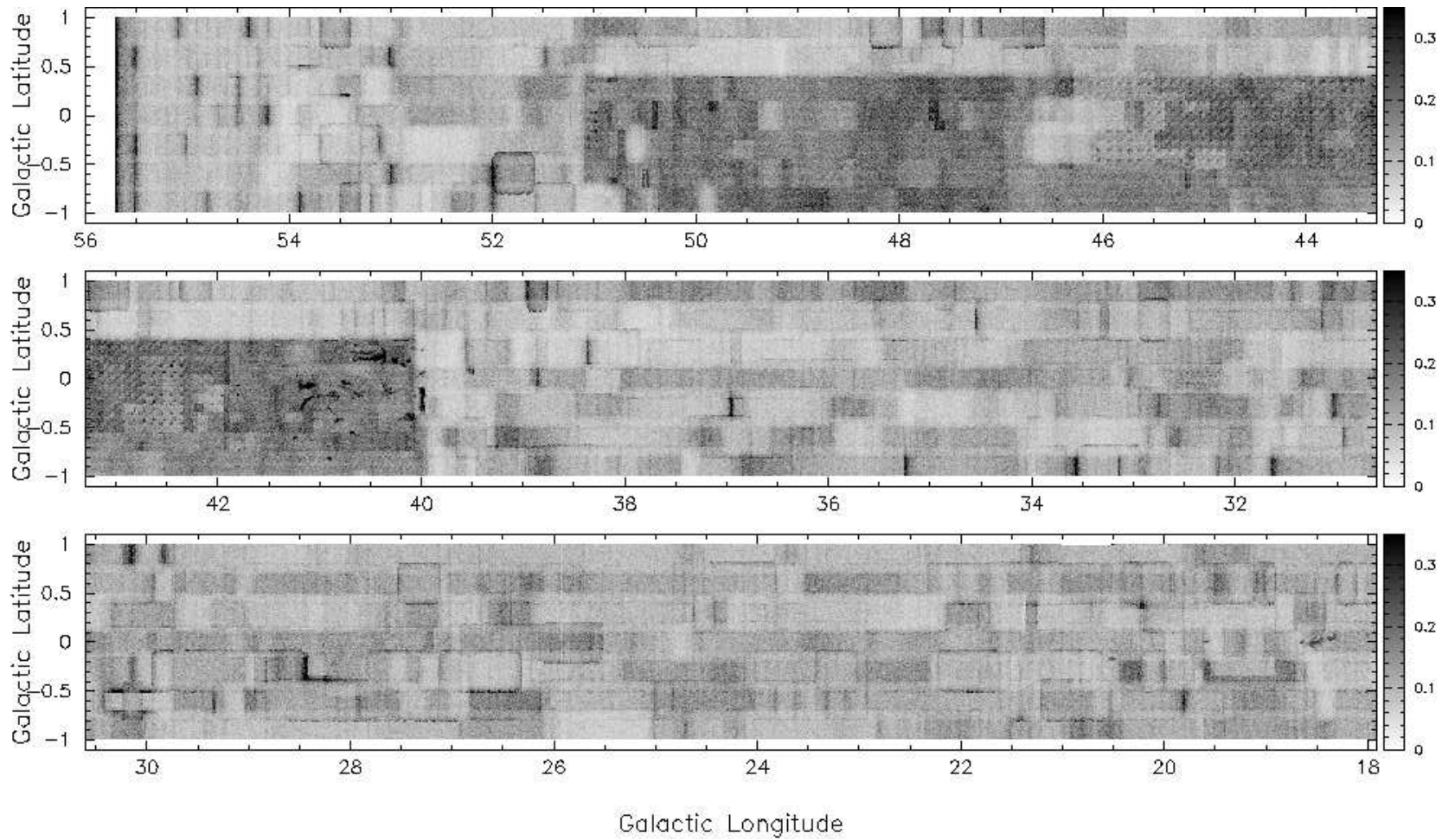}
\caption{\label{noise-image} An image of the rms noise temperature, $\sigma$(\tastar), for the entire GRS. The image
units are in K. The noise temperatures were determined using data from 130 to 135\,\kms\, for
$\ell \leq 40\arcdeg$ and from 80 to 85\,\kms\, for $\ell > 40\arcdeg$.
The compact features are artifacts due to emission at these velocities. The noise
patterns differ for the different observing modes.}
\end{sidewaysfigure}

\begin{figure}
\includegraphics[angle=90,width=0.5\textwidth]{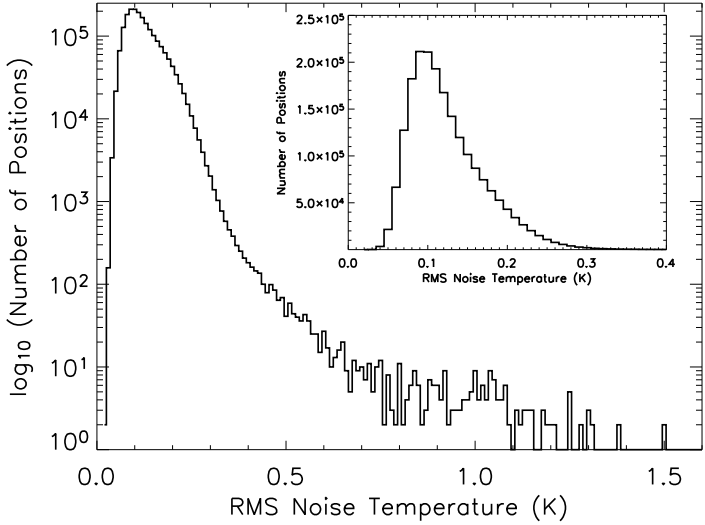}
\caption{\label{noisestats} Histogram of the RMS noise temperature, $\sigma$(\tastar), for all ($\ell,b$)
positions in the GRS. The RMS noise temperatures were determined using data from 130 to 135\,\kms\, for
$\ell \leq 40\arcdeg$ and from 80 to 85\,\kms\, for $\ell > 40\arcdeg$.
The inset shows the same histogram but on a linear
rather than logarithmic scale. The data are binned by 0.01~K intervals.}
\end{figure}

We made a two dimensional image of the rms noise temperature, $\sigma$(\tastar), distribution
throughout the GRS (Fig.~\ref{noise-image}). We plot a histogram of the rms
noise temperature versus the number of ($\ell$,$b$) pixels with that noise
temperature in Figure~\ref{noisestats}.  Once again, the inset shows the same
plot but on a linear rather than logarithmic scale. The rms noise temperatures
were measured using the velocity region 130 to 135\,\kms\, for
Galactic longitudes $\ell \leq 40\arcdeg$ and from 80 to 85\,\kms\, for Galactic longitudes $\ell > 40\arcdeg$
in every spectrum. We choose these velocities because they are usually
emission-free toward all positions. Toward a few positions, however, Galactic
emission is seen. Nevertheless, because this emission occupies a small
fraction of the survey's total solid angle, it will not significantly affect the
noise characteristics. 

The noise distribution shows a sharp peak at $\sigma$(\tastar)~$\sim$~0.1~K. In addition,
a long tail extends to higher noise temperatures. Again, this tail shows that the 
observations were taken under a variety of weather conditions, elevations
and observing modes. Consequently, the noise is not uniform for all positions.
We find that 25\% of the positions have noise temperatures $<$ 0.08 K, 50\% $<$ 0.10 K,
and 75\% $<$ 0.13 K. We will use the latter value as a typical sensitivity for the
GRS. Note that this value compares well to the dispersion of the antenna temperature
$\sigma$ (\tastar) = 0.12 K for all voxels shown in Figure~\ref{emission_stats}.

\begin{figure}
\includegraphics[angle=90,width=0.5\textwidth,clip=true]{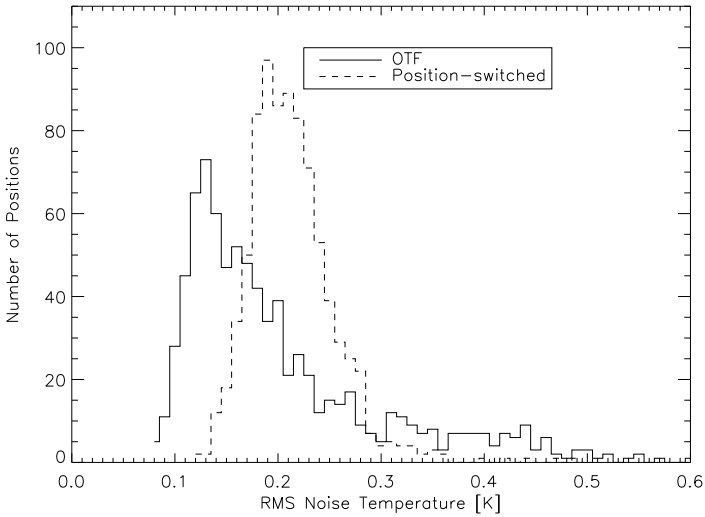}
\caption{\label{noise_stats_compare}Comparison of the RMS noise temperature, $\sigma$(\tastar), for OTF and position-switched
data. For the comparison, we use a small region of the survey ($40\arcdeg\le\ell\le 40\fdg 1$ and 
$-0\fdg 3\le b\le0\arcdeg$) that was observed in both of these modes. The improved sensitivity of the OTF data
(peak at 0.13 K) compared to the position-switched data (peak at 0.20 K) is clearly evident. The data are binned by 0.01~K intervals.}
\end{figure}

To compare the noise characteristics of position-switched versus OTF data we observed a
small region of the survey from $40\arcdeg\le\ell\le 40\fdg 1$ and $-0\fdg 3\le b\le0\arcdeg$ 
in both modes. Figure~\ref{noise_stats_compare} plots the noise temperature in this
region for both the position-switched (dashed histogram) and OTF (solid histogram) data.
The improved sensitivity of the OTF data is evident.  The noise in the OTF data peaks at a $\sigma$(\tastar) = 
0.13\,K. The noise distribution of the position-switched data, however, peaks at a higher noise 
temperature of $\sigma$(\tastar) of 0.20\,K. The improvement of the OTF data over the 
position-switched data stems from the increased redundancy of this observing mode, which
results in a larger effective integration time and reduced systematic noise sources.
Figure~\ref{noise-image} also shows these differences.

\begin{figure}
\includegraphics[width=0.9\textwidth,clip=true]{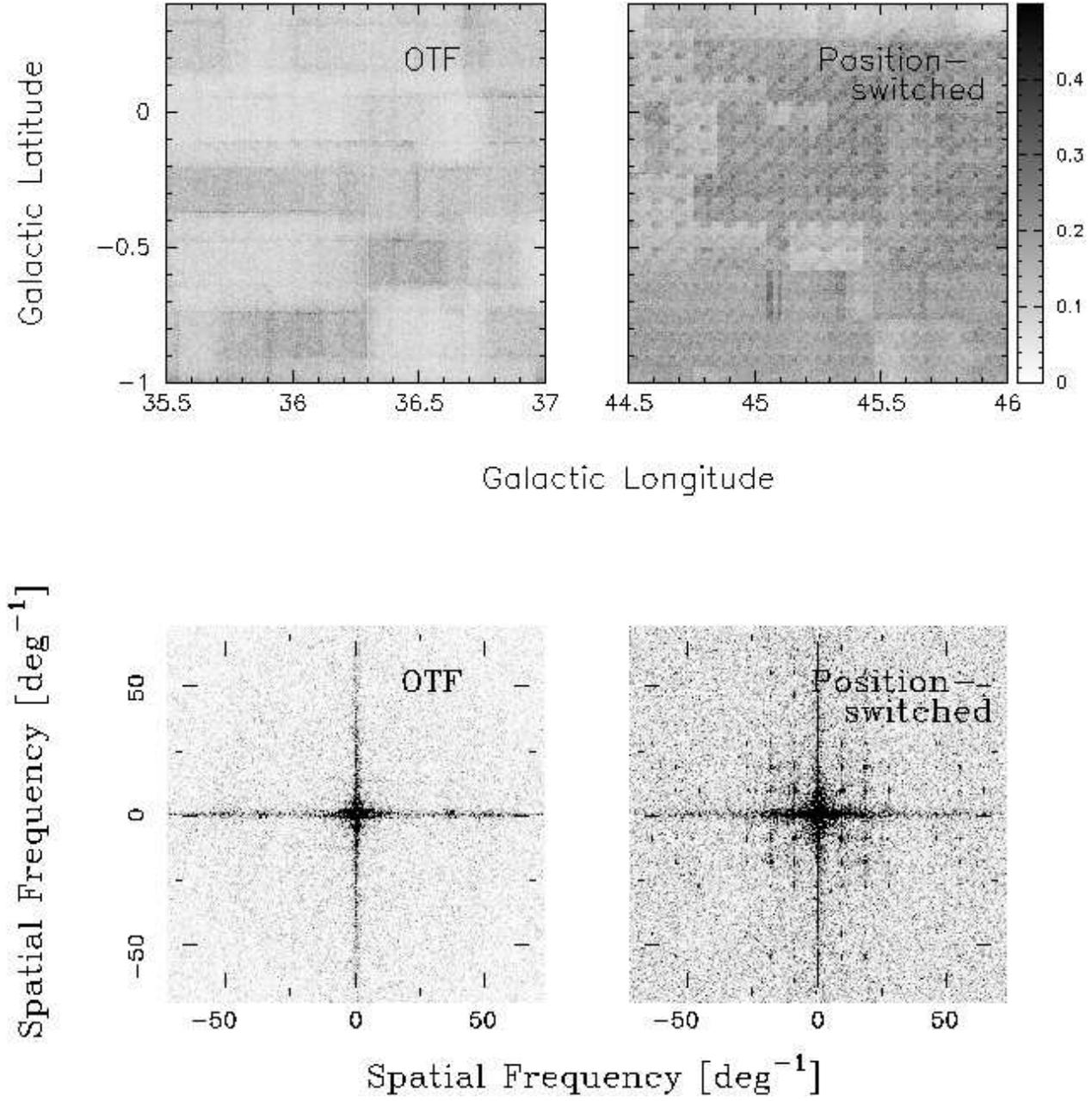}\\
\caption{\label{fft}{\it Upper panel}: Images of the rms noise temperature for small selected regions of the GRS for
the OTF (left) and position-switched modes (right). {\it Lower panel}: Two-dimensional power spectra of the corresponding
noise images.}
\end{figure}

\begin{figure}
\includegraphics[width=0.7\textwidth]{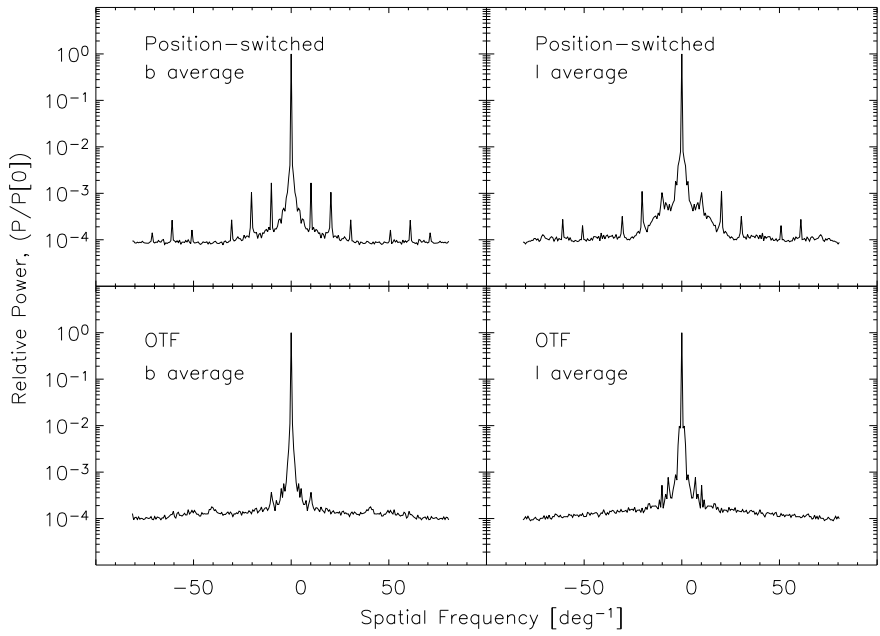}
\caption{\label{fft-cuts} The power spectra of Figure~\ref{fft} averaged
over Galactic latitudes (left) and longitudes (right). Peaks in this
figure represent the correlated noise at certain spatial frequencies.
For the position-switched data, the peaks in both directions 
correspond to multiples of the array size. In contrast, the OTF
data show smaller correlated noise, and also a small peak at
$\sim$ 40 deg$^{-1}$ in the latitude plot due to the sharing
of ``off'' positions in the scanning direction.}
\end{figure}

Because several ``on'' positions share the same ``off'' position, the pixel-to-pixel noise
is not independent but is instead correlated. Thus, spatial averages of the data will
not reduce the noise as the square root of the number of pixels. The degree to which the
noise is correlated depends on the observing mode. In position-switched mode, each position
is observed by only a single pixel and the array is stepped during observations such 
that the same array element observes four adjacent pixels for each ``off'' position. These 
pixels will therefore have correlated noise. Moreover, in the position-switched mode
we step the array by multiples of the array size while mapping. Consequently, noise or gain
patterns across the array will repeat on these spatial scales. Thus, we expect correlated
noise at spatial frequencies corresponding to multiples of the array size. In the OTF mode, however, the array is
scanned such that a single position is observed by multiple pixels. In this mode two adjacent
rows share a common ``off'' position and correlated noise will appear at these spatial
frequencies. Figures~\ref{fft} and \ref{fft-cuts} show these effects. 

In Figure~\ref{fft}, the noise images for small representative regions are shown for both 
OTF and position-switched modes, together with their corresponding power spectra (the square of the two-dimensional
Fourier transform). As expected, the large number of faint, but sharp peaks in the power spectrum for the position-switched
data correspond to the spatial frequencies of multiples of the SEQUOIA array size. In the OTF
data, however, these features are essentially absent. Instead, we see a faint peak in the 
power spectrum at spatial frequencies corresponding to every second row. 

Figure~\ref{fft-cuts} shows the two power spectra of Figure~\ref{fft} averaged over
Galactic longitudes and latitudes. In the position-switched data correlated noise
is seen in both directions because the array was stepped in both Galactic longitude
and latitude. In the OTF data, however, the correlated noise due to shared ``off''
positions, the small peak at spatial frequencies of $\sim$ 40 deg$^{-1}$, 
appears only in the scan direction (Galactic latitude). Moreover,
because OTF data average over several pixels the correlated noise is much smaller.

\section{Data Release}

The GRS data are available at www.bu.edu/galacticring. The data
can be obtained either by requesting a DVD of the complete survey, or
by using our on-line interface. The web page shows the  Figure~\ref{grs-momentmaps} integrated intensity 
image. One may access any specific  FITS data cube by clicking on this image. This action selects
the nearest integrated intensity image and a link to download both the
corresponding FITS data cube and integrated intensity image.

For users interested in smaller regions, or particular molecular complexes, 
we also offer an interface to select any ($\ell$,$b$,v) region within the survey.
One can either enter the required range in Galactic longitude, Galactic latitude, and 
velocity, or a central position in ($\ell$,$b$,v) and a size. Due to file size limitations, the 
maximum Galactic longitude range one can select is 2$\arcdeg$. By default, the full velocity
range for a given region in the survey is returned if no velocity range is
entered. For input regions that overlap two data cubes, the interface
will locate the relevant cubes, extract the region of interest, and generate a new
cube. A link is then provided to download the FITS data cube. We also provide a link to bypass
the interactive process and download individual data cubes by name.
In addition, we offer several simple IDL procedures to aid users with common visualization
and processing tasks. 

\section{Sample images and spectra}

\begin{figure}
\includegraphics[width=\textwidth,clip=true]{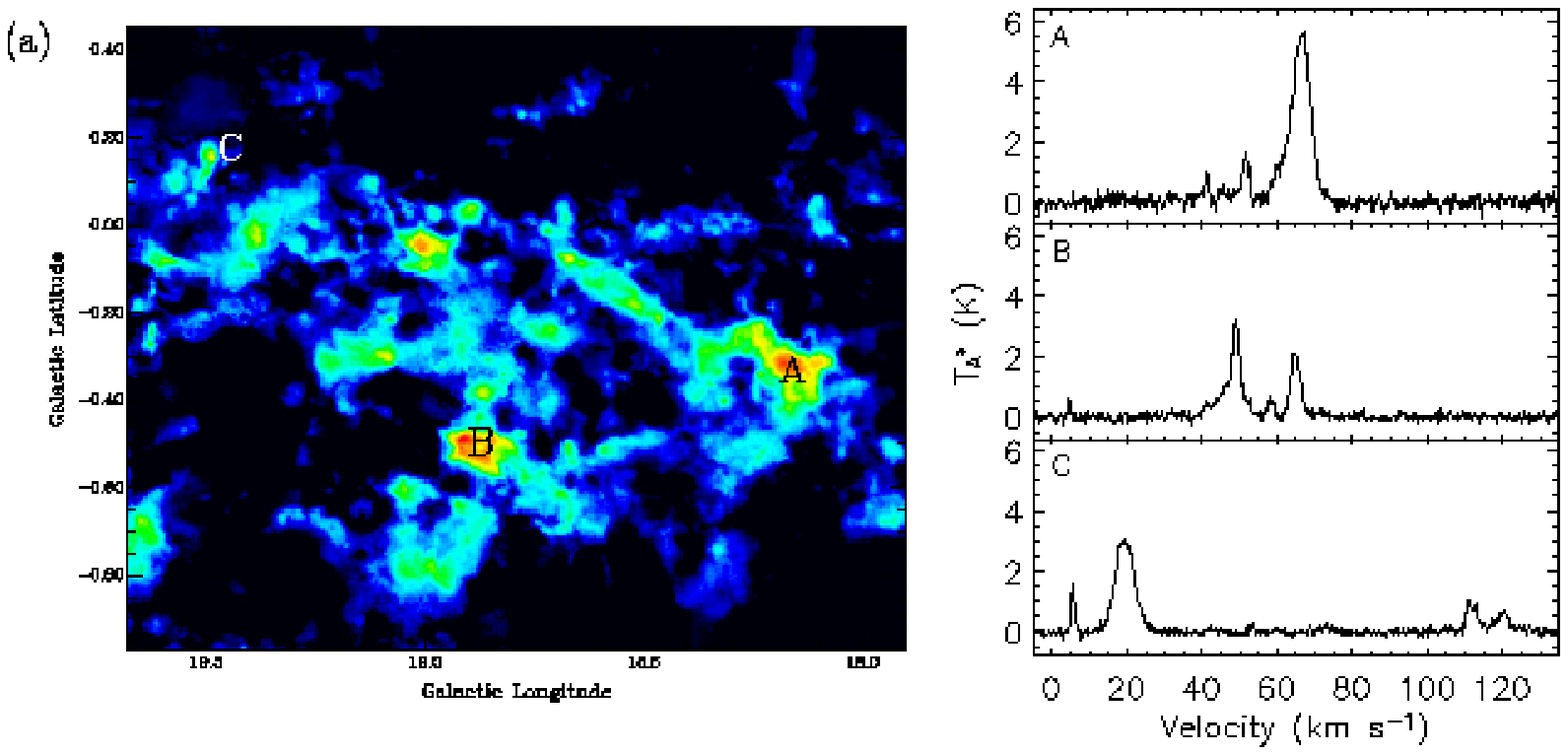}
\includegraphics[width=\textwidth,clip=true]{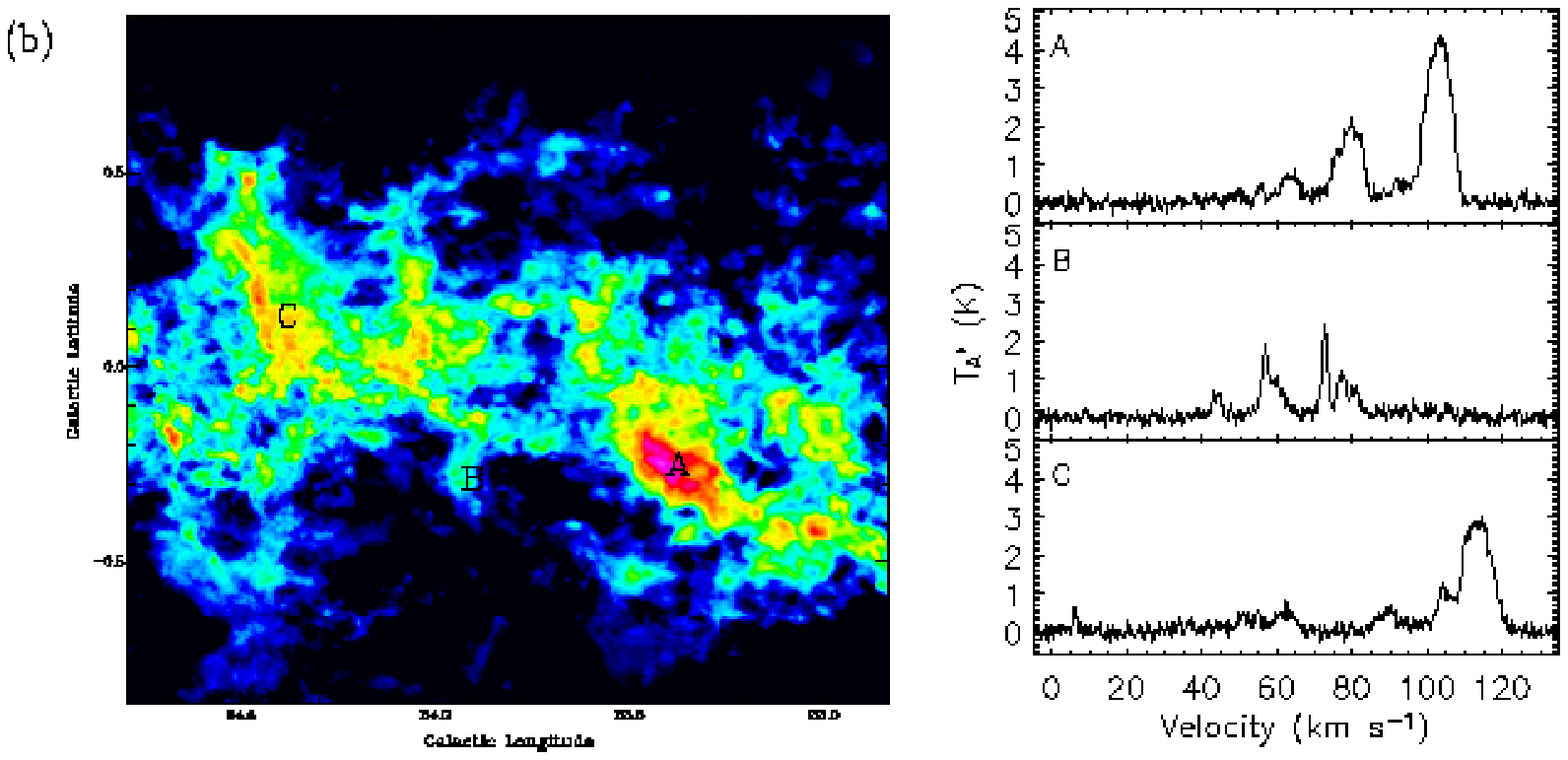}\\
\caption{\label{images-and-spectra}Integrated intensity images and sample spectra toward five
  regions within the GRS. The spectra typically show several distinct velocity
  components, an indication that several distinct molecular clouds lie along the
  line of sight. The location of each spectrum within the
  region is labeled on the image (A, B, and C). The transfer function is such that the following colors
correspond to the following integrated intensities: black (1\,\Kkms), green (20\,\Kkms), red (45\,\Kkms),
and white (100\,\Kkms).}
\end{figure}

The quality of GRS data can be seen in Figure~\ref{images-and-spectra}, which
shows examples of images and spectra toward five distinct regions 
that span a range of total molecular column density.
For each region, we display a moment map and several representative spectra.  The spectra 
typically have several velocity components, which indicate that several distinct
molecular clouds lie along the line of sight.

In Figure~\ref{images-and-spectra}(a), a region at low Galactic longitude is shown.
Since the low longitudes were observed at low elevations,
these data suffer from the highest system temperatures due
to atmospheric extinction. Nevertheless, the moment map and
spectra are not significantly compromised.  Figures~\ref{images-and-spectra}(b) and (c)
show very crowded regions toward the molecular ring.  Figure~\ref{images-and-spectra}(d)
is a region with relatively faint \tcont\, emission (note that there is some faint
\tcont\, emission in the ``off'' beam at a \vlsr\, $\sim$ 12\,\kms) .  Finally,
Figure~\ref{images-and-spectra}(e) shows one of the brightest star-forming regions,
W51.  The images reveal that the GRS has excellent dynamic
range, and the spectra show that the baselines are well
characterized and most of the ``off'' positions are emission-free.

\section{Summary}

Using the SEQUOIA multi pixel array on the FCRAO 14 m telescope, we conducted the
GRS, a new survey of Galactic \tco\, emission. The GRS mapped Galactic longitudes
of $18\arcdeg < \ell < 55\fdg 7$ and Galactic latitudes of $|b| < 1\arcdeg$.
The LSR velocity range surveyed is $-$5 to 135\,\kms\,
for Galactic longitudes $\ell \leq 40\arcdeg$ and  $-$5 to 85\,\kms\, 
for Galactic longitudes $\ell > 40\arcdeg$. The survey achieved better than
half-beam angular sampling (22$\arcsec$ grid; 46$\arcsec$ angular
resolution). 

We have described the telescope and instrumental parameters, the
observing modes, the flux calibration, the data reduction processes, and the 
emission and noise characteristics 
of the dataset. The typical sensitivity of the GRS data is $\sigma$(\tastar) $\sim$ 0.13\,K per
0.21\,\kms\, velocity channel. However, because the survey employed both position-switching and OTF mapping
modes, and observations were conducted under a variety of weather conditions, elevations,
and instrument upgrades, the noise varies across the surveyed region and thus cannot be well characterized
by a single noise temperature. Since survey maps shared common ``off'' positions for several 
``on'' positions, we see correlated noise at certain spatial frequencies. This
correlated noise is more problematic for the position-switching data. 

The GRS data are available to the community at
www.bu.edu/galacticring.  At this website the entire
dataset can be obtained in the form of FITS data-cubes. Users can also 
obtain selected subsets of the data through a simple interface.
The website also contains a few IDL procedures for simple, common processing
tasks.  

The large advance in angular sampling and sensitivity afforded
by SEQUOIA on the FCRAO 14 m telescope makes the GRS an excellent
new dataset for studies of molecular clouds, star forming regions, and
Galactic structure.

\acknowledgments

We gratefully acknowledge observing help from Loren Anderson, Emily Flynn, Ori Fox, 
Dovie Holland, James Kim, Michal Kolpak, Ida Kubiszewski, Casey Law, Kristy McQuinn, 
Mike Martin, Courtney Morris, David Nero, Holly Naylor, Kelsy Rogers, and Todd Veach. 
We also wish to thank the
following people at FCRAO: Peter Schloerb, Neal Erickson, Gopal Narayanan, and Michael Brewer.
The GRS is a joint project of Boston University and Five College Radio Astronomy Observatory, 
funded by the National Science Foundation under grants 
AST 98-00334, AST 00-98562, AST 01-00793, AST 02-28993, and AST 05-07657.

Facilities: \facility{FCRAO}.

\clearpage

\begin{table}
\caption{\label{table-survey-comparison} A comparison of CO surveys of the First Galactic Quadrant.}
\begin{tabular}{lcccc}
\tableline \tableline
Survey                           & GRS         & UMSB$^{1}$  & Bell Labs$^{2}$& Columbia/CfA$^{3}$   \\
\tableline                                                             
Dates                            &  1998--2005 &  1981--1984     &  1978--1992    & 1980--  \\
Transition  (\transition)        &   \tcont    &  \cont          &  \tcont        &\cont     \\
\trstar\, sensitivity per \kms channel (K) & 0.27 & 0.40         &   0.1          & 0.18 \\
Velocity resolution (\kms)       &  0.21       & 1.0             &   0.68         & 0.65      \\  
Longitude coverage (degrees)     & 18 to 55.7  & 8 to 90         &    $-$5 to 117 & 10 to 70   \\
Latitude coverage (degrees)      & $-$1 to 1   & $-$1.05 to 1.05 &  $-$1 to 1     & $-$6 to 6    \\
LSR velocity range (\kms)        & $-$5 to 135 & $-$100 to 200   &  $-$250 to 250 & $-$140 to 140  \\
Angular resolution (arcsec)      & 46          &  45             &   103          & 450       \\
Angular Sampling  (arcsec)       & 22          & 180             &  180           & 225--450         \\ 
Number of Spectra                & 1,993,522   & 40,551          &  73,000        & 54,000     \\   
Total survey region (square degrees)  & 75.4   & 172.2           &  244           & 660        \\ 
\tableline
\end{tabular}
\tablerefs{(1) \cite{Sanders86}, (2) \cite{Lee01}, (3) \cite{Dame01}}
\end{table}

\begin{sidewaysfigure}
\includegraphics[angle=270,width=0.9\textwidth]{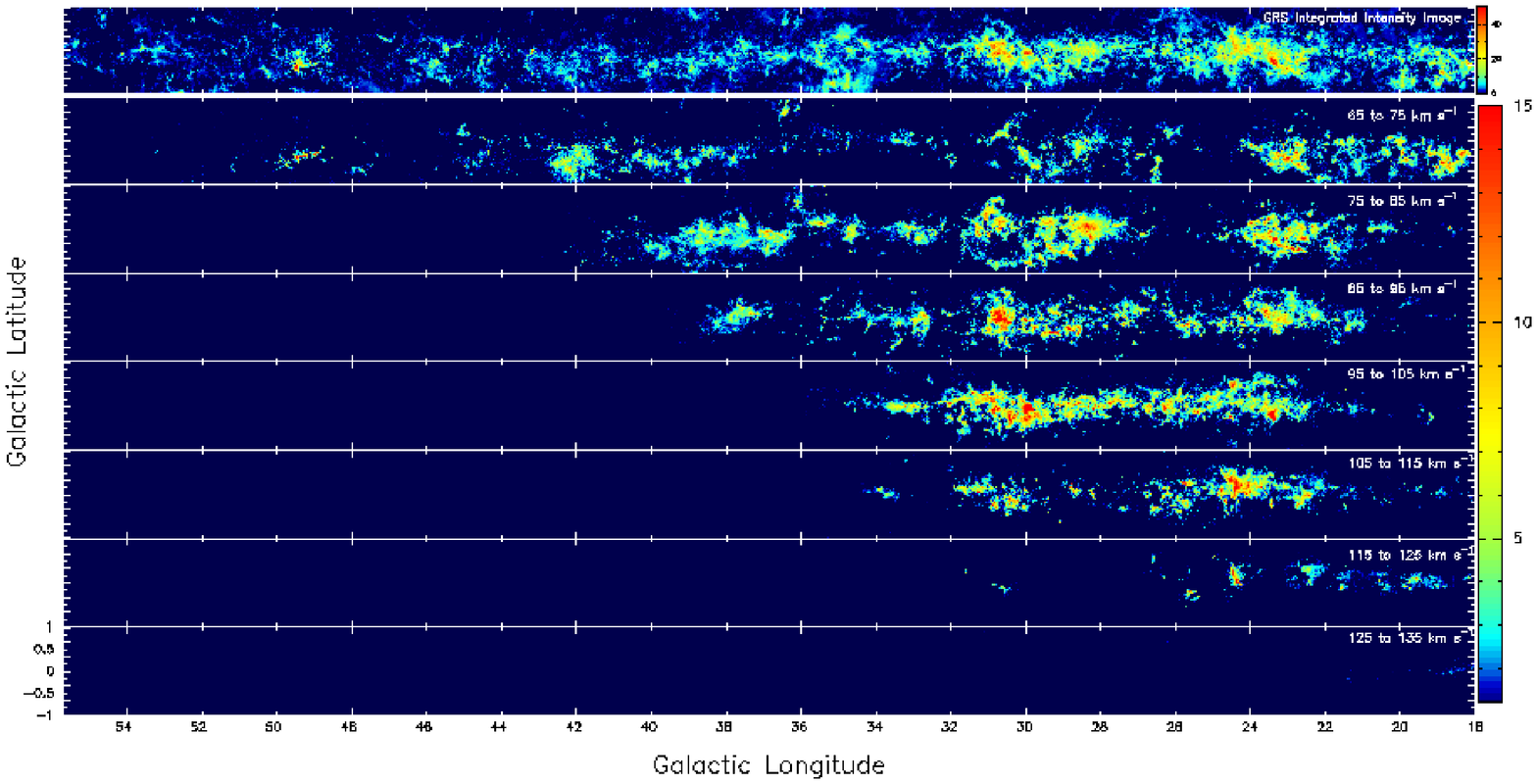}\\
{\small {{Fig.~\thefigure{}.-- \it{continued}}}}
\end{sidewaysfigure}

\clearpage
\begin{figure}
\includegraphics[width=\textwidth,clip=true]{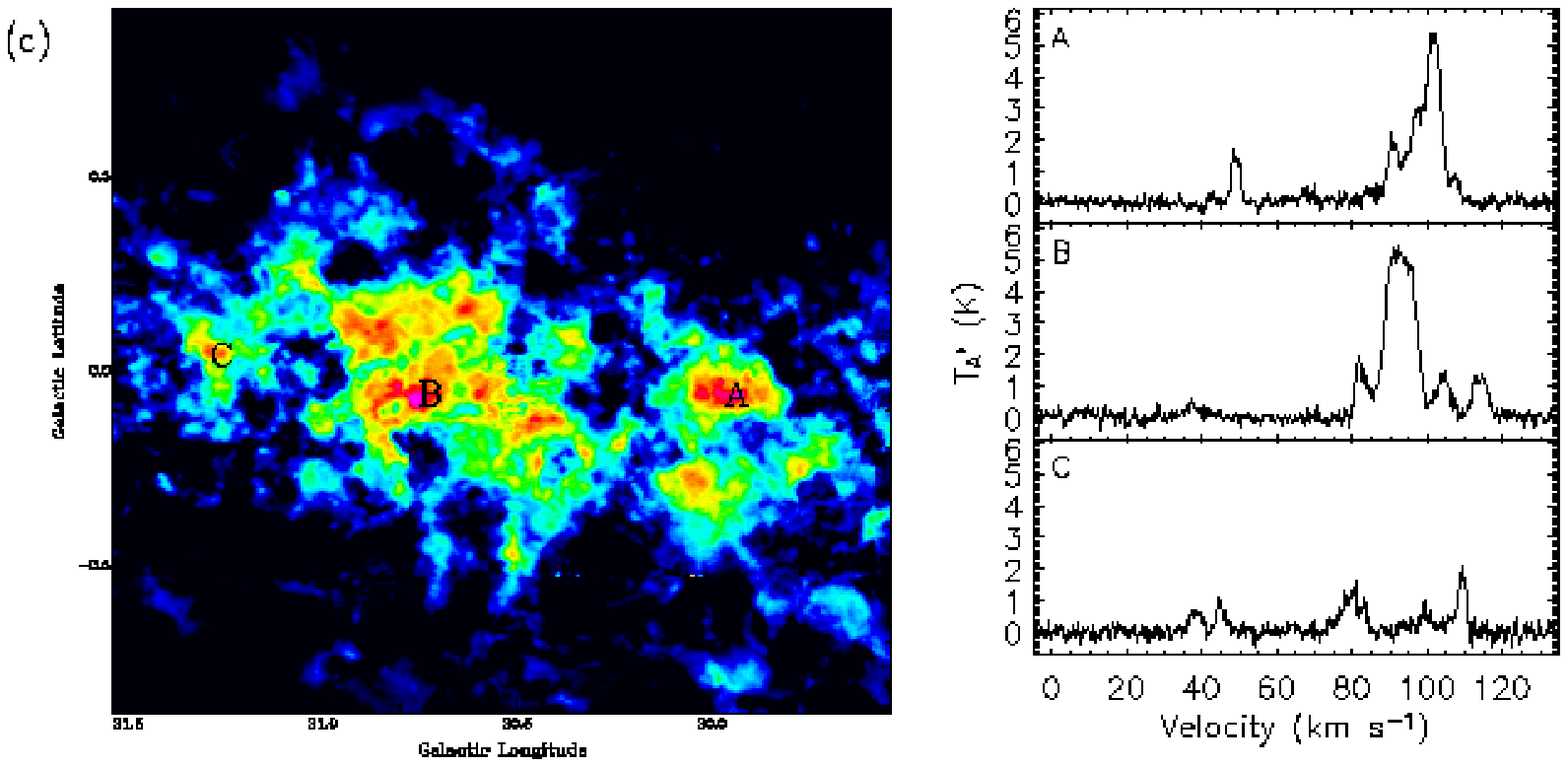} 
\includegraphics[width=\textwidth,clip=true]{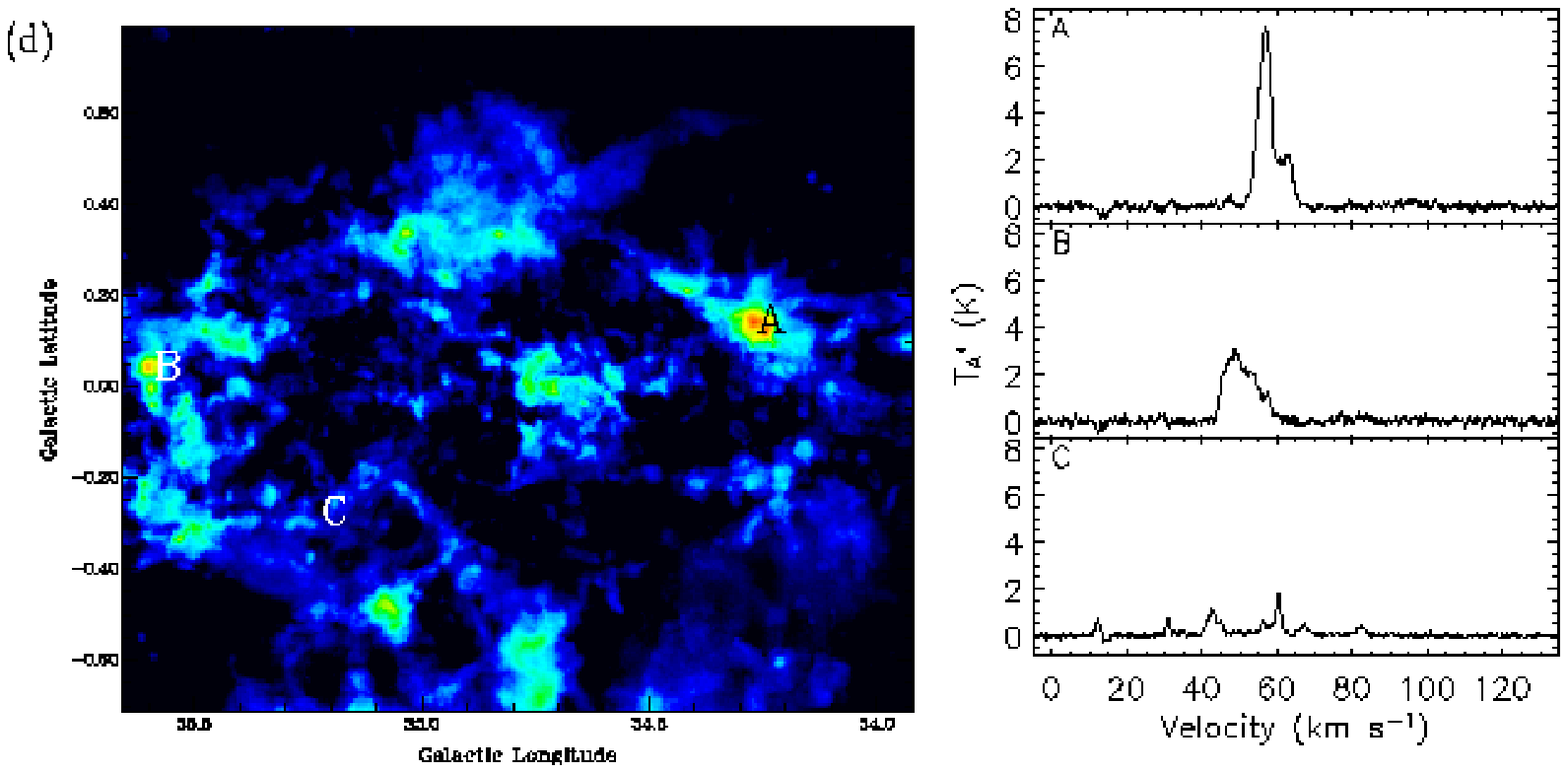}\\
{\small {\sc{Fig.~\thefigure{}.-- \it{continued}}}}
\end{figure}

\clearpage
\begin{figure}
\includegraphics[width=\textwidth,clip=true]{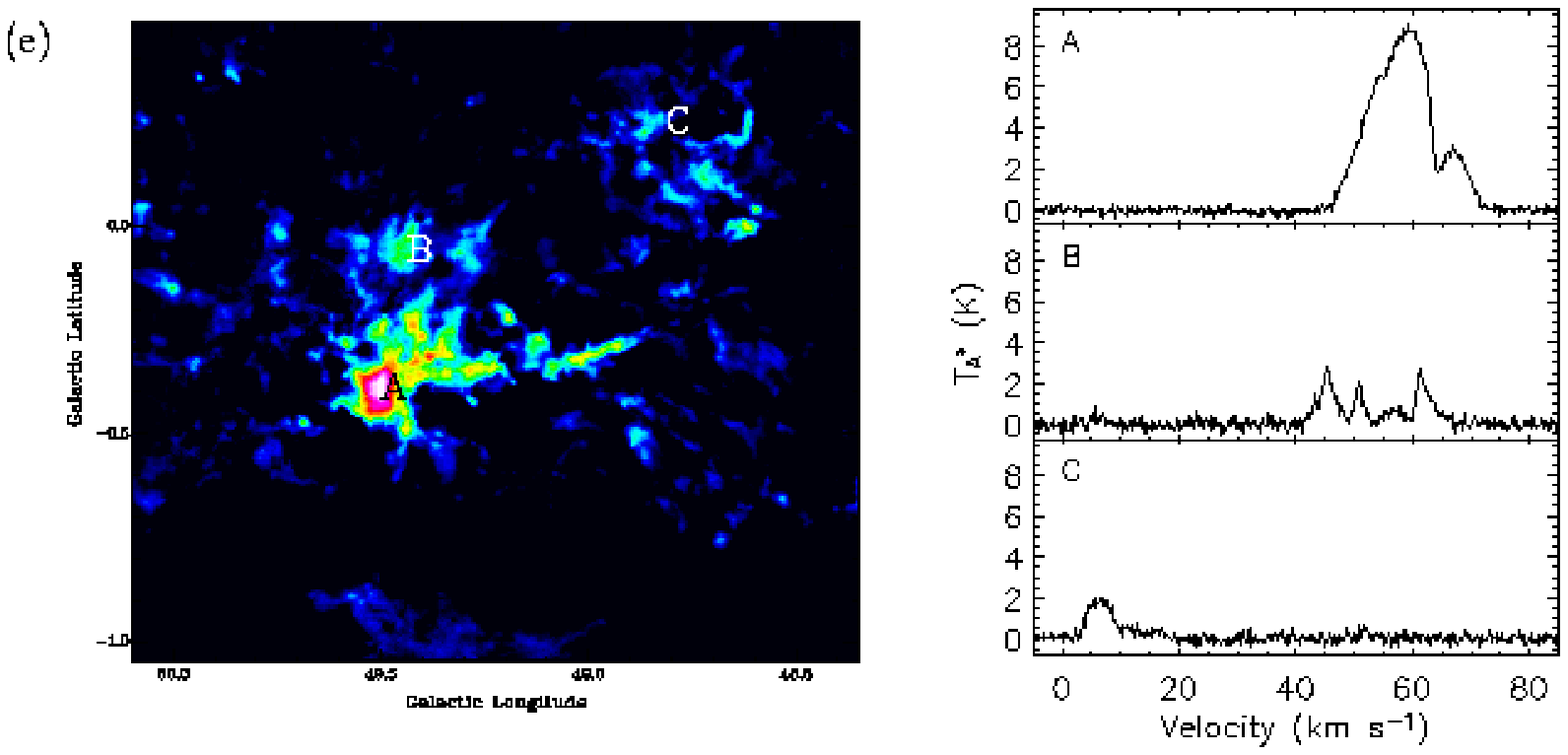}\\
{\small {\sc{Fig.~\thefigure{}.-- \it{continued}}}}
\end{figure}

\end{document}